\documentclass[aps, prl, reprint,twocolumn, preprintnumbers, nofootinbib, showpacs, superscriptaddress]{revtex4}
\usepackage{graphicx, amsmath, amssymb, amsfonts, pifont, bm, cancel}
\usepackage[usenames]{color}
\usepackage{subfigure}
\usepackage[normalem]{ulem} 

\definecolor{MyDarkBlue}{rgb}{0,0,1}

\renewcommand\Re{\operatorname{Re}}
\renewcommand\Im{\operatorname{Im}}

\begin{document}

\leftline{Comment to {\tt mirhosse@optics.rochester.edu}}
\title{Weak-value amplification of the fast-light effect in rubidium vapor}

\author{Mohammad~Mirhosseini}
\email{mirhosse@optics.rochester.edu}
\affiliation{The Institute of Optics, University of Rochester, Rochester, New York 14627, USA}
\author{Gerardo~Viza}
\affiliation{Department of Physics, University of Rochester, Rochester, New York 14627, USA}
\author{Omar~S.~Maga\~na-Loaiza}
\affiliation{The Institute of Optics, University of Rochester, Rochester, New York 14627, USA}
\author{Mehul~Malik}
\affiliation{The Institute of Optics, University of Rochester, Rochester, New York 14627, USA}
\affiliation{Institute for Quantum Optics and Quantum Information (IQOQI), Austrian Academy of Sciences, Boltzmanngasse 3, A-1090 Vienna, Austria}


\author{John~C.~Howell}
\affiliation{The Institute of Optics, University of Rochester, Rochester, New York 14627, USA}
\affiliation{Department of Physics, University of Ottawa, Ottawa ON K1N 6N5, Canada}

\author{Robert~W.~Boyd}
\affiliation{The Institute of Optics, University of Rochester, Rochester, New York 14627, USA}
\affiliation{Department of Physics, University of Ottawa, Ottawa ON K1N 6N5, Canada}

\date{\today}

\begin{abstract}

We use weak-value amplification to enhance the polarization-sensitive fast-light effect from induced Raman absorption in hot rubidium vapor. We experimentally demonstrate that projecting the output signal into an appropriate polarization state enables a pulse advancement of 4.2 $\mu$s, which is 15 times larger than that naturally caused by dispersion. More significantly, we show that combining weak-value amplification with the dispersive response of an atomic system provides a clear advantage in terms of the maximum pulse advancement achievable for a given value of loss. This technique has potential applications for designing novel quantum-information-processing gates and optical buffers for telecommunication systems.
\end{abstract}

\maketitle

Due to Kramers-Kronig relations, a sharp change in the absorption or the transmission of an optical medium results in a large modification of the group index \cite{Boyd:2003wk}. Controlling the group velocity using slow and fast-light is an enabling technology with many applications in photonics \cite{Camacho:2009cd,Reim:2010ik,Shi2007b,Boyd:2002ud}. Additionally, fast-light provides a unique testbed for studying the fundamental physics behind superluminal pulse propagation. While slow light can be achieved with no appreciable loss using electromagnetically induced transparency (EIT) \cite{Fleischhauer:2005da}, one needs to operate close to the center of an absorption line to achieve fast-light \cite{Mikhailov:2004es}. In this situation, the large amount of absorption limits both the applications and the magnitude of the observed fast-light effect.  

A weak measurement is a generalized form of quantum measurements, in which a weak unitary interaction is followed by a strong projective measurement \cite{Aharonov:1988fk, Dressel:2014ks}. Unlike the standard measurements, the result of a weak measurement, known as a weak value, can be beyond the range of eigenvalues of the measured operator \cite{Dressel:2010ub}. This property, known as weak-value amplification (WVA), has been used before to sensitively measure a variety of effects, such as a transverse beam deflection \cite{Dixon2009,Ritchie1991,Hosten:2008ih, MaganaLoaiza:2014kf}, phase \cite{Brunner:2010dt}, velocity \cite{Viza:2013kq}, and time delay \cite{Brunner:2004cf}. Further, it has been suggested that the weak-value amplification can be used to enhance nonlinear optical effects in the few photons regime \cite{Feizpour:2011bs}. 

In this work, we amplify the negative time delay associated with the ``superluminal" pulse propagation of an optical pulse in hot rubidium vapor. The fast-light effect is caused by an induced Raman absorption profile of the rubidium hyperfine structure in a pump-probe nonlinear interaction. Due to this effect, the polarization of an optical pulse is weakly coupled with its arrival time. By appropriately preparing and post-selecting the polarization states of the pulse, we can effectively engineer the dispersion properties of the medium \cite{Brunner:2004cf, Solli2004}, and thus amplify the weak coupling between polarization and arrival time. Using this technique, we were able to advance the peak of an optical pulse by an amount that is up to 15 times larger than the original fast-light advancement.

The fast-light effect due to absorption and the enhancement due to weak-value amplification are both lossy processes. Here, we study how the achieved temporal advancement scales as a function of loss due to atomic absorption and compare it to the scaling as a function of loss due to post-selection in WVA. Remarkably, we find that for a given value of loss, an optimized combination of both these processes provides a larger time advance than that obtained by just increasing the atomic absorption itself. In light of the ongoing debate on the usefulness of WVA \cite{Ferrie:2014gf, Jordan:2014jv, Knee:2014dd}, we find this result to be both timely and significant.

\begin{figure}[t]
\includegraphics[width=8cm]{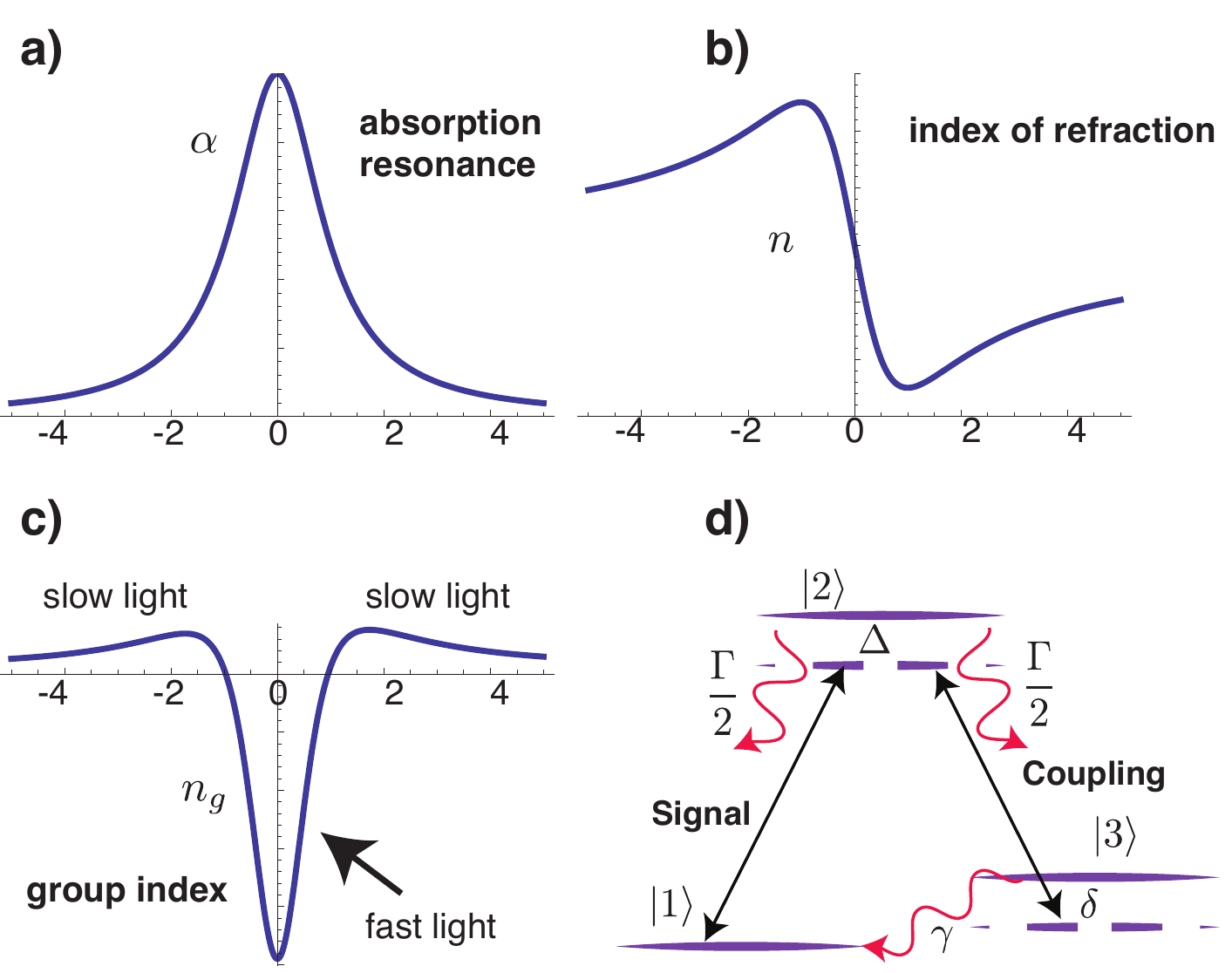}
  \caption{ a) The absorption profile of a Lorentzian lineshape. b) The refractive index associated with the absorption line can be calculated using the Kramers-Kronig relations. c) The group index for the same line. d) Schematic diagram of a three-level Lambda system.}
  \label{fig:Fig_KK}
\end{figure}

\emph{Tunable group delay from atomic response.} In a dispersive medium the group velocity and the phase velocity are not the same. The group velocity and the group index can be calculated from the standard results
\begin{equation}
v_g = c/n_g, \qquad n_g = n + \omega \frac{dn}{d\omega}.
\end{equation}
Slow and fast-light correspond to the situations where $n_g >> 1$ and $n_g <1$, respectively. Due to the Kramers-Kronig relation, a sharp change in the absorption coefficient can lead to a substantial change in the group index. A large pulse advance in fast light can be achieved by operating in a wavelength close to the center of an absorption line (See Fig. \ref{fig:Fig_KK}). However, the large amount of loss in this region limits the amount of maximum negative delay that can be achieved in practice.

We use a nonlinear-process to induce a polarization-sensitive absorption line in an atomic vapor. Consider a three-level atomic $\Lambda$ system, where levels 1 and 2 are connected via the signal field $\frac{\Omega_p}{2} e^{-i(\omega_p t-kz)}+ c.c$ and level 2 and 3 are connected by a strong coupling field $\frac{\Omega_c}{2} e^{-i(\omega_c t-kz)}+ c.c$. The detunings are defined as $\Delta = \Delta_p = (\omega_2 - \omega_1)-\omega_p$, $\Delta_c = (\omega_2 - \omega_3)-\omega_c$ and $\delta = \Delta_p - \Delta_c$. In the case where the coupling and the signal have orthogonal polarizations the susceptibility at the signal frequency can be calculated as 
\begin{equation}
\chi (\Delta,\delta,\Omega_c) = \beta \frac{\delta - i \gamma}{(\delta - i \gamma)(\Delta - i {\Gamma}/{2})- {{|\Omega_c|}^2}/4 }. 
\end{equation}

Here, $\gamma$ and $\Gamma$ are the excited-state and the ground-state decoherence rates, respectively. The factor $\beta$ is equal to $N \mu^2/\hbar \epsilon_0$, where $N$ is the number density and $\mu$ is the transition dipole moment between levels 1 and 2. This formula can explain many interesting results including the EIT effect. For large single photon detuning values, $\Delta >> \Gamma$, this expression can be approximated by a Lorentzian line shape. In this case the susceptibility can be written as
\begin{eqnarray}\label{Eq:Lorentzian}
\chi (\Delta>>\Gamma) = \beta \frac{{|\Omega_c|}^2}{4\Delta^2}\frac{\delta'+ i \gamma'}{{\delta'}^2 + {\gamma'}^2},
\end{eqnarray}
In the above $\delta' = \delta- \delta_0$, $\gamma' = \gamma + \gamma_0$, where $\delta_0 = \frac{{|\Omega_c|}^2\Delta}{4\Delta^2 + \Gamma^2}$ and $\gamma_0 = \frac{{|\Omega_c|}^2\Gamma}{8\Delta^2 + 2\Gamma^2}$.

It should be emphasized that in this configuration, the polarization component of the signal that is orthogonal to the polarization of the signal beam experiences an advancement in time with respect to propagation in vacuum, whereas the component of the signal polarized parallel to the coupling beam experiences no modification in the group delay while traveling through the medium. When the signal field is comprised of both polarization components, the differential time advance can be enhanced by a projective measurement in the polarization \cite{Brunner:2004cf, Solli2004}. It is well known that the amplification of a time delay obtained in this manner is in fact an interference effect that can be fully understood using classical electromagnetic theory. However, expressing this phenomenon within the weak-value formalism leads to a simpler and more elegant description that is easier to understand.

\begin{figure*}[t]
\includegraphics[width=0.8\textwidth]{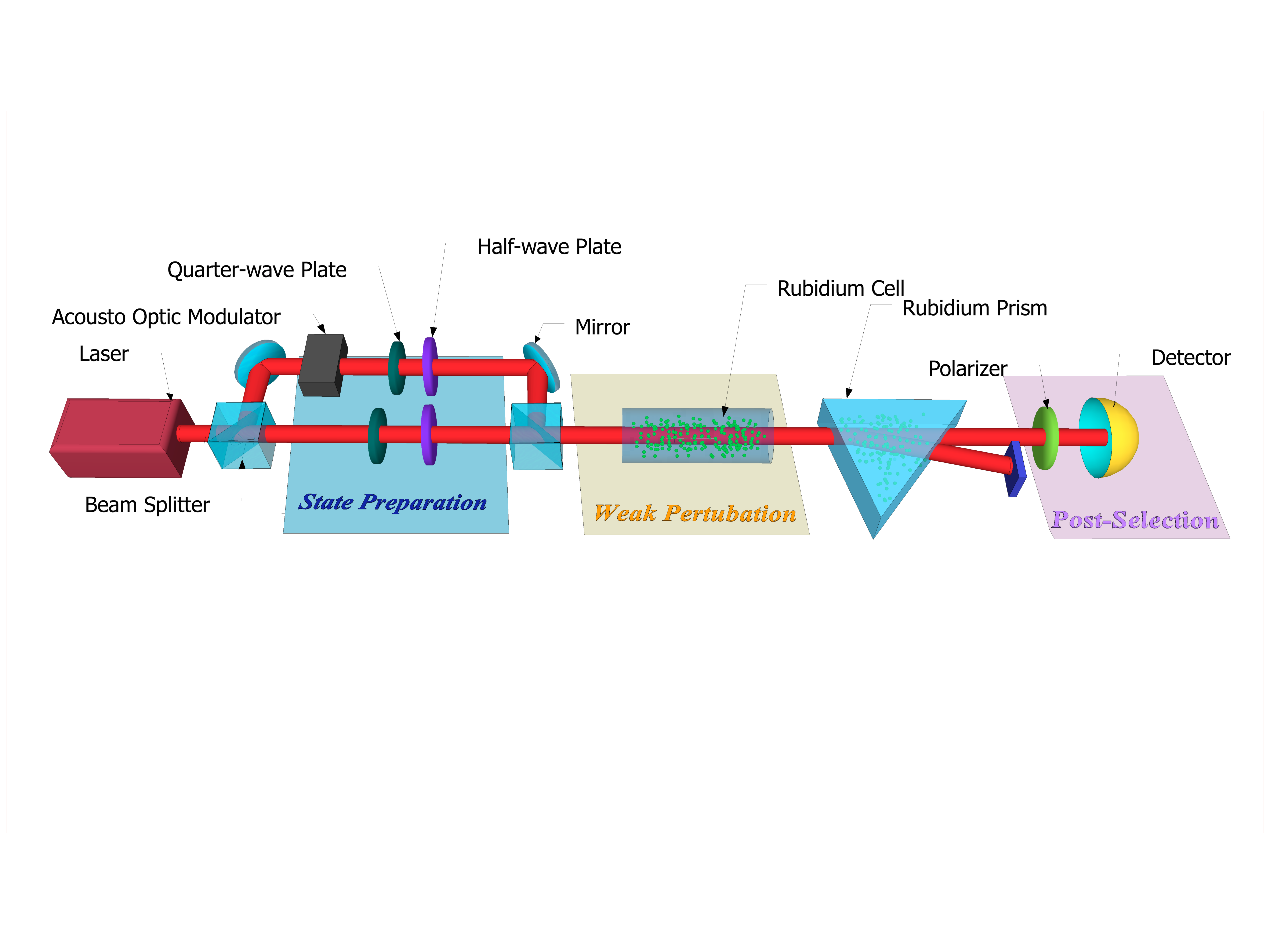}
  \caption{{Schematic of the experimental setup.} The continuous wave laser beam is divided to two copies using a non-polarizing beam splitter. The probe is frequency shifted using an acousto-optic modulators. The polarization state of the pump and the probe are controlled using wave plates. The post-selection is done using a polarizer and a fast detector. The output photo-current is analyzed by an oscilloscope.}
  \label{fig:Schematics}
\end{figure*}

\emph{Weak-value amplification.} We cast the propagation of an optical pulse through the atomic vapor in the language of quantum state measurement. The polarization and the temporal state of the signal beam before the cell can be described as
\begin{equation}
|\Psi_{in} \rangle = \frac{1}{\sqrt{(1+T)}}\left( |H \rangle + \sqrt{T} |V \rangle\right)\otimes |{f}(t)\rangle e^{-i\omega_0t},
\end{equation}
where we have assumed a quasi-monochromatic single optical mode with a pulse shape described by \(|{f}(t)\rangle\). The horizontal and vertical polarization states are shown as \(|H \rangle\) and \(|V \rangle\) respectively. Since the polarization component \(|H \rangle\) attenuates upon propagation, it is weighted by a larger factor to pre-compensate for the effect of loss. The power transmission efficiency of propagation through the cell for the horizontally polarized beam is denoted by $T$.
We now consider a case where the coupling beam is polarized in the vertical direction. In this situation, the state of the signal beam in the output can be described as \cite{Fleischhauer:2005da,Vudyasetu:2010um}
\begin{equation}
|\Psi_{out} \rangle = \sqrt{\frac{T}{(1+T)}}\left(|H \rangle |{f}(t+{t_0})\rangle  + |V \rangle |{f}(t)\rangle \right)e^{-i\omega_0t}.
\end{equation}
Here, $t_0 = - (n_g -1)\frac{L}{c}$ is the absolute value of the group delay for a propagation length $L$. We have dropped the common delay time between the two polarization states in order to simplify the notation. It is seen that horizontal polarization experiences attenuation and an advancement in time compared to the vertical component of the field.  Additionally, we have assumed the optical path length in the medium for the two polarization components are equal. This results in the convenient phase difference of zero between the two polarization components in the output. In practice, a non-zero phase difference can always be pre-compensated by changing the polarization state of the input signal beam.

We use the weak value formalism for the case where the advancement time $t_0$ is much smaller than the temporal duration of the pulse $f(t)$. Although our formalism closely follows that of Ref.\,\cite{Brunner:2004cf}, the large amount of time advance and the tunability provided by the atomic system in our experiment offers a degree of control absent from previous realizations. Furthermore, we provide a comparison of the loss vs. time advance from WVA to the one obtained from the KK relations. This analysis provides theoretical evidence for the efficacy of combining weak value amplification from the natural dispersive time advance from atomic response.

Assuming a Gaussian pulse shape, a post-selection in the a linear polarization state with an angle $\theta$ with respect to the frame of the experiment results in  

\begin{equation}
\mid\Psi_{PS} \rangle \approx \sqrt{\frac{T}{(1+T)}} \left(\cos\theta |H \rangle   + \sin\theta |V \rangle\right)\otimes |{f}(t+{A_{w}t_0})\rangle e^{-i\omega_0t}.
\end{equation}
Here, $\Psi_{PS}$ is the polarization and the temporal state of the post-selected beam. 

The weak value \(A_{w}\) corresponds to the temporal shift of the optical pulse and can be calculated using the formula
\begin{equation}
A_{w} = \frac{\langle\Phi_{\theta}\mid\hat{A}\mid\Psi_{in}\rangle}{\langle\Phi_{\theta}\mid\Psi_{in}\rangle} = \frac{\cos \theta }{\sin \theta +\cos \theta }, 
\end{equation}

where \(|\Phi_{\theta}\rangle\) is the post-selection polarization state and the measurement operator is  \(\hat{A} = |H\rangle\langle H|\). It can be seen in Fig.~3 that choosing the post-selection angle $\theta$ close to $45^\circ$ results in a large amplification factor $A_{w}$. More interestingly, it is possible to achieve a negative amplification and hence convert a time delay to a time advance and vice versa. 

\begin{figure}[h!]
  \centerline{\includegraphics[width=8.0cm]{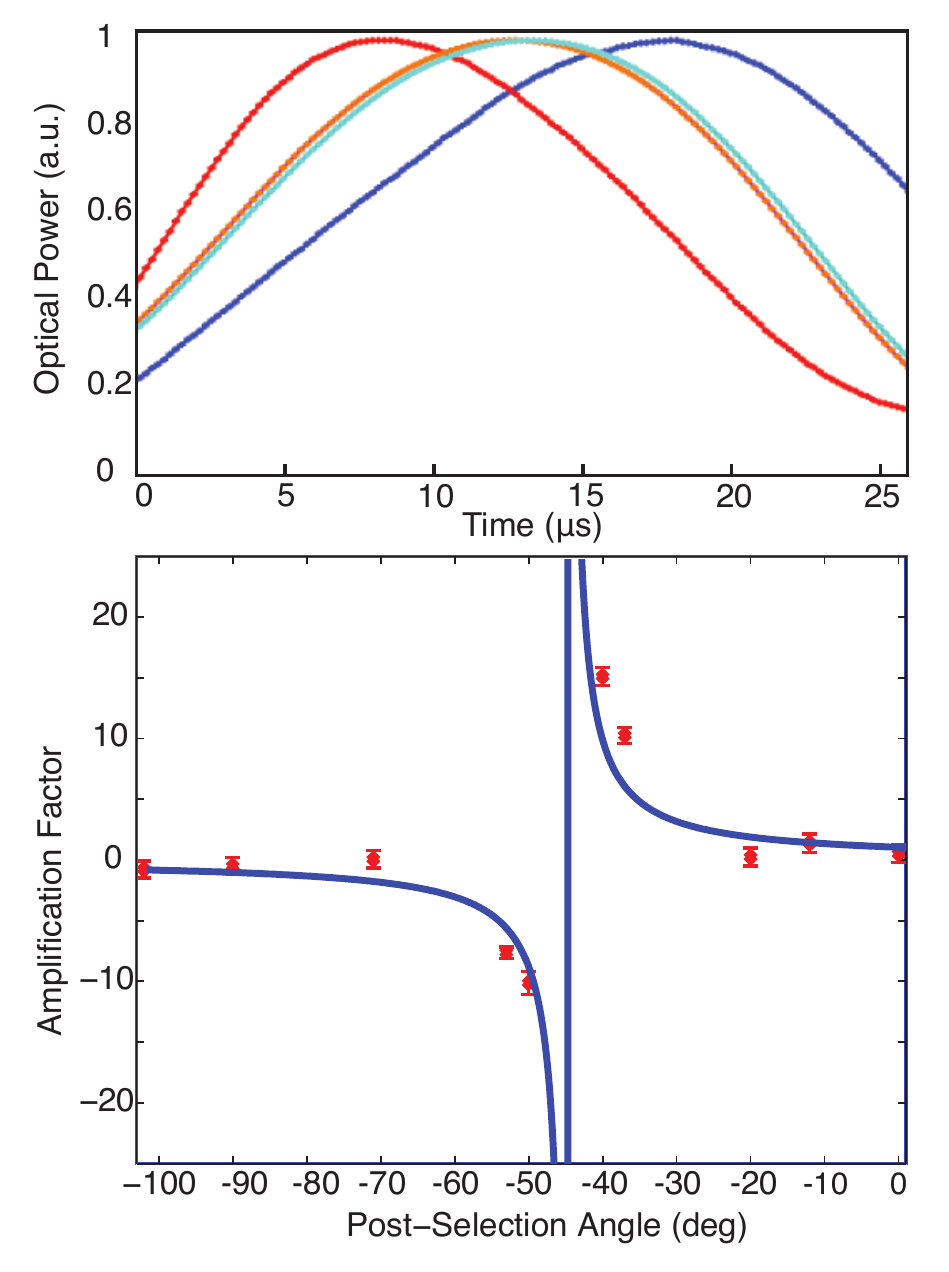}}
  \caption{Top: Measured optical power as a function of time. The cyan and orange curves show the detected optical power for the $|V\rangle$ and $|H\rangle$ sates. The red and blue curve correspond to post-selection angles $\theta = -40^\circ$ and $\theta = -50^\circ$. Bottom: The amplification factor as a function of post-selection angle. The amplification factor is calculated from data by performing a Gaussian fit.The error bars correspond to the 95 $\%$ confidence interval.}
  \label{fig:Fig_Pulse}
\end{figure}

\emph{Experiment.} 
We realize the Raman absorption profile using warm atomic rubidium vapor. A sketch of the experimental setup is depicted in Fig. \ref{fig:Schematics}. The beam from a 795-nm narrow-line-width tunable diode laser is passed through a tapered-fiber amplifier to obtain a high power beam. The frequency of the laser is tuned to have the coupling beam near resonant with the $|2\rangle$ to $|3\rangle$ transition. The signal beam is obtained by frequency-shifting part of the the laser beam by 3.035 GHz by double-passing it through a tunable acousto-optic modulator. An 8 cm vapor cell is heated using strip heaters inside a teflon tube enclosed by antireflection-coated windows at each end to achieve temperature stability. The cell is shielded from stray magnetic field by a Mu-metal tubing. The vapor cell contains both rubidium isotopes in their natural abundance. In addition, we also have 20 Torr neon in the cell, which acts as a buffer gas. The temperature of each vapor cell is about $80^\circ \text{C}$, resulting in a number density of about $10^{12} \text{cm}^{-3}$.

The coupling and the signal beams are aligned co-linearly to achieve the absorption dip. We use an atomic prism to filter out the coupling beam after the cell \cite{Starling:2012wr}. This prism contains isotopically pure 87Rb and is heated to $100^\circ \text{C}$ to achieve a large dispersion ($dn/d\lambda$). The coupling and the signal beams propagate at different angles once they exit the prism. We collect the signal beam and project it on an arbitrary superposition of the two beams using a polarizer. The pulses are detected using a fast photo-detector diode.

We have chosen to work in the center of the absorption line ($\delta' = 0$) to achieve a maximal time advance $t_0 = 0.28\,\mu\text{s}$. The top part in Fig.\ref{fig:Fig_Pulse} presents measured optical power as a function of time for 4 different post-selection angles. It can be seen that the post-selection can greatly modify the amount of delay, and even flip the sign of it results for two different post-selection angles. We have calculated the amplification factor, by fitting a Gaussian wave form to the pulse and measuring the position of its center. The value of amplification factor as a function of post-selection angle from the experiment is plotted in the bottom part of Fig.\ref{fig:Fig_Pulse}. It can be seen that the delay gets drastically amplified as \(\theta\) approaches \(-45^\circ\).

\emph{The scaling of loss and group-delay.}
The analysis and the experimental results above suggest that the WVA can be combined with the dispersive response of an atomic system to provide extra control over the value of group delay. However, to get an appreciable amplification factor one needs to post-select on a state that is nearly orthogonal to the input state. In this situation, the efficiency of the process is significantly reduced. This is, in fact, a universal property associated with weak values and the usefulness of WVA in presence of this additional loss has been a topic of debate recently \cite{Ferrie:2014gf, Jordan:2014jv, Knee:2014dd}.

Before analyzing the effect of loss from post-selection, we investigate the relation between loss and group delay from the atomic response. For operation at the center of the absorption line, $\delta' = 0$, we have (see the appendix)

\begin{equation}
T = \exp{(-2\gamma'  t_0)}.
\end{equation}
In deriving this relation we have used Eq.\,\ref{Eq:Lorentzian} and the approximation $n \approx 1+ \frac{1}{2}\chi$. It is evident that increasing the absolute value of time advance via increasing the non-linear interaction results in an exponential decrease in the transmission efficiency. Consequently, the maximum achievable delay from atomic response for a given value of minimum acceptable transmission efficiency can be calculated as

\begin{equation}\label{Eq.Loss}
t_\text{atom} = -\frac{\ln{T}}{2\gamma'}.
\end{equation}
Equation \ref{Eq.Loss} calculates the maximum amount of time advance that can be achieved from the atomic response for a given value of transmission efficiency. As we showed earlier, the WVA amplification can be employed to increase the amount of group delay at the cost of a further increase in the loss. For a given value of loss, the maximum value of achievable time advance can be calculated by optimizing the absorption from the atomic response and the loss from post-selection. Performing the analysis we find the maximum value of time advance as (see the appendix)
\begin{equation}\label{Eq.Loss2}
{t_\text{WVA}} =  \frac{1}{2\gamma'}\max_\theta\left(\frac{\cos \theta }{\sin \theta +\cos \theta }\ln{\left[\frac{2\sin^2{\left(\theta+\frac{\pi}{4}\right)}}{T}-1\right]}\right).
\end{equation}

\begin{figure}[h]
  \centerline{\includegraphics[width=8.0cm]{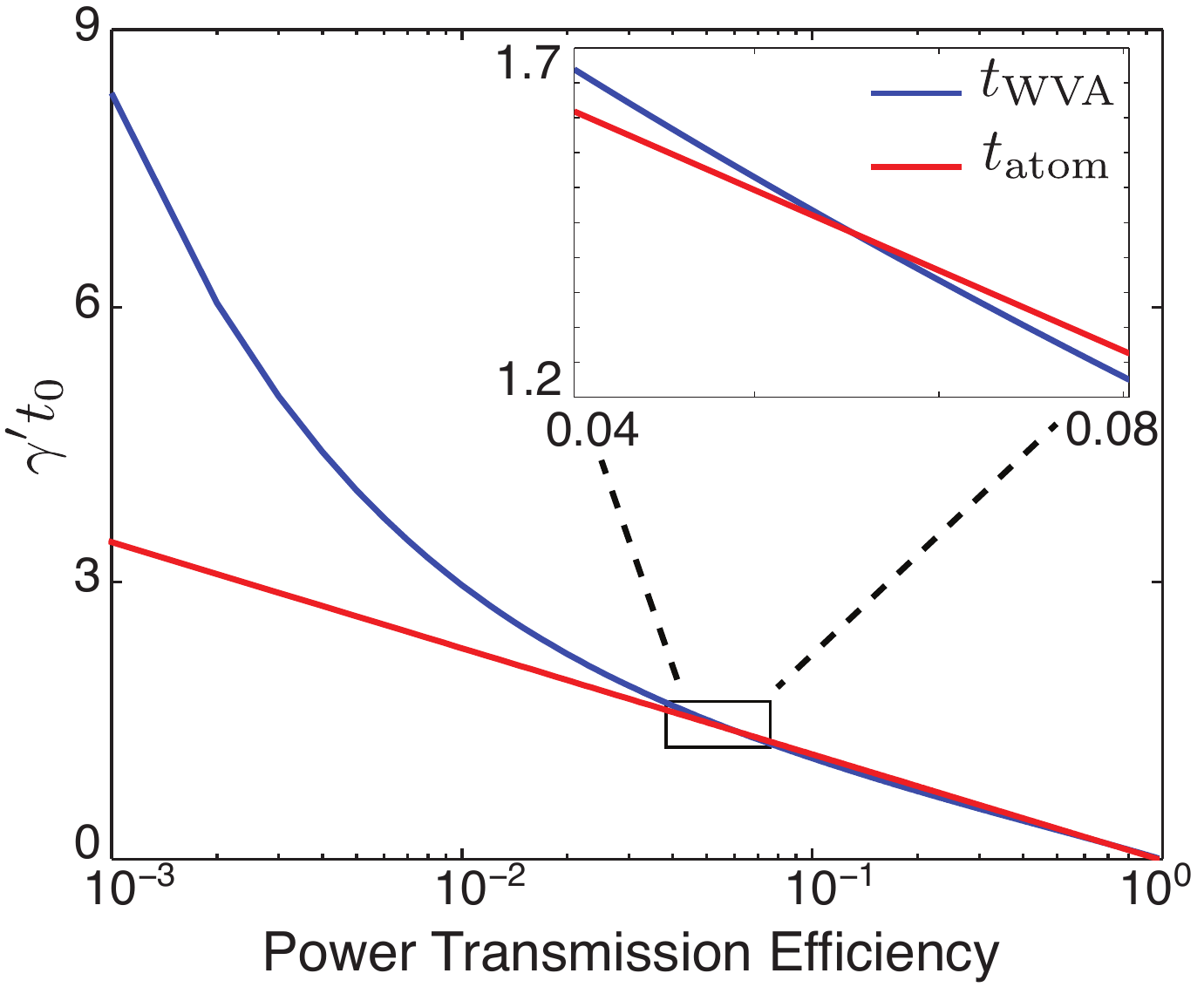}}
  \caption{{Maximum achievable time advance as a function of transmission efficiency (calculated from theory)}}
  \label{fig:LossScaling}
\end{figure}
The solutions of Eq.~\ref{Eq.Loss2} have been calculated numerically and plotted in Fig.~\ref{fig:LossScaling}, along with the solutions for Eq.~\ref{Eq.Loss}. It is evident that the WVA procedure provides a slightly lower time advance for large values of transmission efficiency. However, as the transmission efficiency decreases, the time advance obtained by using WVA grows rapidly, crossing the advancement obtained from the atomic response at $T \approx 5\%$. For all values of $T$ lower than this value, WVA provides a larger time advance than that obtained from the atomic response alone. This showcases a clear instance where WVA provides an advantage for the estimation of a small interaction parameter.

\emph{Conclusions} We have used weak-value amplification to enhance the fast-light effect caused by electromagnetically-induced absorption in warm Rubidium vapor. By appropriately preparing and post-selecting the polarization state of an optical pulse, we have obtained an advancement in time that is 15 times larger than that obtained from the atomic resonance. The enhancement from WVA can also be tuned to convert a time advance into a time delay and vice versa. Additionally, we have shown that when the total transmission through the system is lower than 5\%, the use of WVA provides a clear enhancement in the amount of time advance possible. Our technique has the potential to allow better control over photonic technologies that use fast and slow light, such as quantum-information-processing gates and optical buffers for telecommunication systems.

We gratefully acknowledge valuable discussions with S. Gillmer, J. Vornehm, Z. Shi, J. Ellis and D. Gauthier. This work was supported by the DARPA/DSO InPho program, the Canadian Excellence Research Chair (CERC) program, Army Research Office grant number W911NF-12-1-0263. OSML acknowledges the support from CONACYT and the Mexican Secretar\'ia de Educaci\'on P\'ublica (SEP).

\bibliography{TDA}{}
\section{Appendix}
Using the susceptibility in Eq.\,\ref{Eq:Lorentzian} and assuming \mbox{$n \approx 1+ \frac{1}{2}\chi$} we get
\begin{equation}
n = \Re{[n]} + i \Im{[n]}= 1+ \beta \frac{{|\Omega_c|}^2}{8\Delta^2}\frac{\delta'+ i \gamma'}{{\delta'}^2 + {\gamma'}^2}.
\end{equation}
The group delay can be calculated using the real part of the group index. Assuming $\Delta>>\Gamma$, and $\Delta>>|\Omega_c|$  we have
\begin{equation}
\left.{n_g}\right\vert_{\delta' = 0} = 1+ \beta \frac{{|\Omega_c|}^2}{8\Delta^2}\frac{\omega}{\gamma'^2}.
\end{equation}
The group delay can be calculated using the group index as $\frac{Ln_g}{c} $. The value of group delay includes the propagation time in vacuum. The differential time advance can be found as
\begin{equation}
t_0 = \beta\frac{L}{c} \frac{{|\Omega_c|}^2}{8\Delta^2}\frac{\omega}{\gamma'^2}
\end{equation}
We find the value of absorption at the center of the Lorentzian line as
\begin{equation}
\alpha = \left.\frac{\omega}{c} \Im{[n]} \right\vert_{\delta' = 0} = \frac{\beta}{8c} \frac{{|\Omega_c|}^2}{\Delta^2}\frac{\omega}{\gamma'}.
\end{equation}
Subsequently, the power transmission efficiency is
\begin{equation}
T = \exp{\left(-2 \alpha L\right)} = \exp{\left( -2\gamma't_0 \right)}.
\end{equation}
The relation above can be inverted to find the maximum achievable time advance for a minimum acceptable transmission efficiency 
\begin{equation}
t_\text{atom} = -\frac{\ln{T}}{2\gamma'}.
\end{equation}
The alternative strategy is to initially achieve a time advance $\tilde{t}_0$ from the atomic response, at the cost of a reduction in the transmission efficiency for one of the polarization components to $\tilde{T}$. The input beam is described by the state
\begin{equation}
|\Psi_{in} \rangle = \frac{1}{\sqrt{1+\tilde{T}}}\left( |H \rangle + \sqrt{\tilde{T}} |V \rangle\right)\otimes |{f}(t)\rangle e^{-i\omega_0t},
\end{equation}
and the state after the post-selection 
\begin{equation}
\mid\Psi_{PS} \rangle = \sqrt{\frac{\tilde{T}}{1+\tilde{T}}} \left(\cos\theta |H \rangle + \sin\theta |V \rangle\right)\otimes |{f}(t+{A_{w}\tilde{t}_0})\rangle e^{-i\omega_0t}.
\end{equation}
As a result, the total transmission efficiency is 
\begin{equation}
T = {\left|\langle \Psi_{PS} |\Psi_{PS} \rangle\right|}^2 = {\frac{\tilde{2T}}{1+\tilde{T}}} \sin^2{\left(\theta+ \frac{\pi}{4}\right)}
\end{equation}
The total time advance is equal to 
\begin{equation}
{t_0} = A_{w}\tilde{t}_0 = \frac{\cos \theta }{\sin \theta +\cos \theta } \tilde{t}_0.
\end{equation}
Using the relation $\tilde{t}_0 = -\frac{\ln{\tilde{T}}}{2\gamma'}$ we have
\begin{equation}
{t_0}(\theta) =  \frac{1}{2\gamma'}\frac{\cos \theta }{\sin \theta +\cos \theta }\ln{\left[\frac{2\sin^2{\left(\theta+\frac{\pi}{4}\right)}}{T}-1\right]}.
\end{equation}
And finally, we maximize the time advance to find the optimal post-selection angle
\begin{equation}
{t_\text{WVA}} =  \frac{1}{2\gamma'}\max_\theta\left(\frac{\cos \theta }{\sin \theta +\cos \theta }\ln{\left[\frac{2\sin^2{\left(\theta+\frac{\pi}{4}\right)}}{T}-1\right]}\right).
\end{equation}
\end{document}